\documentclass{aa}
\usepackage{psfig}
\def\msun{M_{\odot}}
\newcommand{\mc}{\multicolumn}

\def\degmark{^\circ}
\def \rsun {\ifmmode$R$_{\odot}\else R$_{\odot}$\fi}
\def \nh {N${\rm _H}$}
\def \ngal {N${\rm _{gal}}$}
\def \hcm {\hbox {\ifmmode $ H atoms cm$^{-2}\else H atoms cm$^{-2}$\fi}}
\def\approxgt{\mathrel{\hbox{\rlap{\lower.55ex \hbox {$\sim$}}
        \kern-.3em \raise.4ex \hbox{$>$}}}}
\def\approxlt{\mathrel{\hbox{\rlap{\lower.55ex \hbox {$\sim$}}
        \kern-.3em \raise.4ex \hbox{$<$}}}}
\newcommand {\sax} {{BeppoSAX}}

\begin{document}

\thesaurus{(13.01.1; 13.25.3)}

\title{The absorption properties of gamma-ray burst
afterglows measured by BeppoSAX}

\author{Alan Owens\inst{1} \and
M. Guainazzi\inst{1} \and 
T. Oosterbroek\inst{1} \and 
A. Orr\inst{1} \and 
A.N. Parmar\inst{1} \and
E. Costa\inst{2} \and 
M. Feroci\inst{2} \and
L.~Piro\inst{2} \and
P. Soffitta\inst{2} \and
D. Dal Fiume\inst{3} \and
F. Frontera\inst{3,4} \and
E. Palazzi\inst{3} \and
E. Pian\inst{3} \and
J. Heise\inst{5} \and
J.J.M. in 't Zand\inst{5} \and
M.C.~Maccarone\inst{6} \and
L. Nicastro\inst{6}
}

\institute{
Astrophysics Division, Space Science Department of ESA,
ESTEC, 2200 AG Noordwijk, The Netherlands
\and
Instituto Astrofisica Spaziale, CNR., Via Fosso del Cavaliere, 
I-00131 Roma, Italy
\and
ITESRE/CNR, via Gobetti 101, 40129 Bologna, Italy
\and
Dipartimento di Fisica, Universita di Ferrara, 
Via Paradiso 12, 44100 Ferrara, Italy
\and
Space Research Organization Netherlands, Sorbonnelaan 2,
3584 CA Utrecht, The Netherlands
\and
IFCAI/CNR, Via U. La Malfa 153, I-90146, Palermo, Italy
}

\date{Received ; accepted }

\maketitle

\begin{abstract}
We present an analysis of the X-ray absorption
properties of 6 gamma-ray burst (GRB)
afterglows measured with BeppoSAX. Between 
8 hrs and 20 hrs after the initial GRB trigger,  
individual spectra can be described by a power-law with 
a photon index of $\sim$2 and 
absorption, \nh, marginally consistent with the galactic value. 
Taken collectively, the data are inconsistent with zero \nh\ at 
the $>$99.999\% confidence level.
The data are only marginally
consistent with a distribution of column densities varying 
as the total galactic \nh\ in the direction of each of the 
bursts ($\chi^2$=9.6 for 6 degrees of freedom). 
The data are consistent with cosmological models in which GRB occur 
within host galaxies. 
By simultaneously fitting a power-law spectral model with
\nh\ fixed at the galactic value and additional, redshifted,
absorption to all 6 afterglow spectra, 
the best-fit average \nh\ within the host objects 
is found to be (1.01${\pm ^{0.28}_{0.51}}$) 
$\times$ 10$^{22}$ atom cm$^{-2}$. This value is compatible with the 
host galaxy column densities inferred from optical data for GRB970508, 
GRB971214 and GRB980329, supporting the hypothesis that GRB occur within  
heavily absorbed star forming regions of their host galaxies. 
\end{abstract}


\section{Introduction}

The BeppoSAX observation of gamma-ray
burst (GRB) afterglows at X-ray wavelengths has 
ushered in a new era of gamma-ray burst studies. The breakthrough has 
been achieved using a conventional GRB detector in conjunction with 
a wide field X-ray camera and narrow field X-ray telescopes. The GRB 
detector verifies the nature of the burst and the wide field camera accurately 
determines its position, allowing a suite of highly sensitive, but 
narrow field instruments, to be slewed to the burst position within 
a matter of hours. The rapid dissemination of arc-minute 
sized burst coordinates has resulted in the successful detection of 
afterglows at other wavelengths.
Although the data are still preliminary, we may use this new information 
to investigate the burst progenitor, mechanism and burst site.

\subsection{Burst astronomy with BeppoSAX}
 
The narrow field instruments (NFI) on the Italian-Dutch 
satellite BeppoSAX (Boella et al. 1997a) have approximately 1$\degmark$
fields of view and include the imaging low- and 
medium energy concentrator spectrometers (LECS, 0.1--10~keV,
Parmar et al. 1997; and MECS, 2--10~keV, Boella et al. 1997b). 
The NFI are co-aligned and are normally operated 
simultaneously. In addition, the payload includes two wide field cameras 
(WFC, 2--30~keV, Jager et al. 1997) which observe in directions 
perpendicular to the NFI and a gamma-ray burst monitor (GRBM, 40--700~keV,
Feroci et al. 1997). These last two instruments 
allow the detection of X-ray transient phenomena and gamma-ray
bursts.

After a GRBM trigger, WFC data are analyzed ${post-facto}$ for a 
simultaneous
X-ray event and, if one is found, a Target of Opportunity (TOO) declared. 
The burst location is quickly derived to an accuracy of a few arcminutes 
and the NFIs slewed to this position. The whole process 
takes $\sim$8 hours. For some 
bursts several TOOs were scheduled, typically a few 10$^4$~s 
long, occurring $\sim$0.3 days, 
days and several days after the initial trigger. 
For positive detections, the LECS and MECS uncertainty radii
are typically 1$'$. To date, there have been 12 such detections --
GRB970111 (Feroci et al. 1998a), GRB970228 (Costa et al. 
1997; Frontera et al. 1998), GRB970402 (Nicastro et al. 1998a), 
GRB970508 (Piro et al. 1998), GRB971214 (Antonelli et al. 1997), GRB971227 
(Piro et al. 1997a), GRB980329 (in 't Zand et al. 
1998), GRB970425 (Pian et al. 1998), GRB980515 (Feroci et al. 1998b), 
GRB980519 (Nicastro et al. 1998b), GRB980613 (Costa et al. 1998) and 
GRB980703 (Galama et al. 1998a) and of these, only 7 are statistically 
sufficient for spectral 
analysis. The analysis that follows
is limited to the first 6 bursts which are listed in Table 1.
For completeness the table also
includes a 7th burst 
(GRB970828) which, while not observed by 
BeppoSAX, was measured spectroscopically by ASCA (Yoshida et al. 1998).

All observed X-ray afterglows decay with 
time as $\sim$${\rm t^{-1.3}}$ and their spectra are well 
fit by an absorbed power-law 
of photon index, $\alpha \sim$2. 
Some show outbursts 
of activity (Piro et al. 1998) and some evidence of spectral evolution 
(Piro et al. 1997b; Yoshida et al. 1998). By far the most interesting 
events are those for which an optical transient has also been detected. 
The additional optical information allows the X-ray data to be 
placed in context. For example, it is found that the spectral and 
temporal properties at optical wavelengths mirror those at X-ray 
wavelengths, ostensibly in agreement with the predictions of even 
the simplest fireball model. In addition, red-shifted emission and absorption 
features have been detected in the spectra of 3 optical afterglows 
(Metzger et al. 1997a; Kulkarni et al. 1998; Djorgovski et al. 1998b).
The derived red-shifts are 0.835 
(GRB970508), 3.42 (GRB971214) and 0.966 (GRB980703). In the cases 
of GRB971214 
(Kulkarni et al. 1998), GRB980329 (Djorgovski et al. 1998a) and most 
recently GRB980703 (Djorgovski et al. 1998b), 
host galaxies have also been detected and probably in the 
case of GRB970228 (Sahu et al. 1997; Fruchter et al. 1998).

\section{Data Analysis}

For each of the BeppoSAX afterglows listed in Table~1,
events were extracted within radii of 8$'$ (LECS) and 2$'$ 
(MECS) of the best-fit source centroids. In cases
where several TOO were made, the analysis 
was limited 
to the first TOO in order to maximize 
the signal-to-noise ratio and
to minimize possible errors induced by the effects of 
spectral evolution, such as those reported for GRB970828 (Yoshida 
et al. 1998) and GRB970508 (Piro et al. 1997b). LECS and MECS 
spectral files, along with their associated responses and 
background files were simultaneously 
fit, keeping fitted parameters tied in both instruments. Because these  
events are weak, great care was taken in selecting blank sky 
regions for background subtraction. For all events, the 0.25, 0.75 and 1.5 keV 
count rates in the LECS standard background file were checked against the 
equivalent rates in the ROSAT all-sky background maps (Snowden et al. 1997). 
In one case (GRB970402), the background at the source was found
to have a significantly different shape than 
the standard background (Nicastro et al. 1998a).
Therefore, in this case the background was extracted from the 
same image at a position diametrically opposite from the source. 
In comparison with the standard background field, the 
differences in fitted spectral parameters were well within errors.

The LECS data were fit over the energy range 0.1--10~keV and MECS 
between 1.8--10.0~keV. It is known from inter-instrument spectral 
calibrations that there can be a position dependent error of 
$\approxlt$20\% in the relative flux 
normalizations between the LECS and MECS. This factor was 
included as a fixed multiplicative parameter during joint spectral 
fitting of LECS/MECS data.
For all afterglows, an absorbed power-law gives the best fit to the 
individual spectra, with  $\chi^2$ typically $\sim$1 per degree of freedom 
(dof). (We note that an unabsorbed black-body also gives almost 
equally good fits.)
In the case of 
GRB970228, only data from the MECS3 unit was included in the fit, 
since the position 
of the burst is close to 
the window support rings of the MECS1 and MECS1 units (see Boella et al. 
1997b). 
The best-fit parameters for each burst are listed in Table~1.

\begin{figure}
\centerline{\psfig{figure=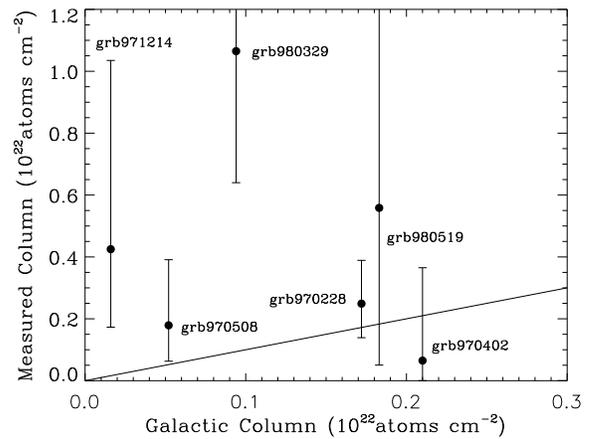,width=8.5cm,angle=0,bbllx=41pt,bblly=267pt,bburx=530pt,bbury=621pt}}
\caption[]{The measured \nh\ for each of the afterglows considered in
the text as a function of \ngal\ in the 
direction of the afterglow. The solid line shows the case if the 
measured columns are equal to the galactic values}
\end{figure}
 
\section{Results}

From Table 1 it is apparent that the fitted photon indices 
are consistent with a constant mean value of 2.21 $\pm$ 0.20. 
This supports a blastwave induced 
synchrotron or inverse Compton interpretation of afterglows
(Waxman 1997; Meszaros \& Rees 1997; Vietri 1997).
In Fig.~1, the fitted hydrogen column density, \nh, for each burst 
is plotted against the galactic column density, \ngal, in that 
direction interpolated from the survey maps of 
Dickey \& Lockman (1990). The estimated precision of the \ngal\ values
is a few times 10$^{20}$ cm$^{-2}$.
Taken collectively, the derived \nh\ are 
inconsistent with zero column density, the $\chi^2$ being 17.5 
for 6 dof $-$ thus excluding all but the most contrived of local 
models. The data are also marginally inconsistent with that of the total 
galactic column density 
along the line of sight of each burst ($\chi^2$ = 9.6 for 6 dof). 
Assuming, that bursts are 
cosmological in origin, we next tested whether the data could support
an additional average absorption above \ngal\
(i.e., intrinsic to the source and/or host object). 
However, at cosmological distances the effective \nh\ is 
increased by (1+$z$)$^{2.6}$ (Morrison \& McCammon 1983) since 
the spectral turnover is reduced 
by (1+$z$) due to the red-shift. Therefore, we 
simultaneously included both the galactic absorption and an
additional red-shifted absorption to represent
intrinsic absorption within the environment local to the afterglow.
In the model convention within XSPEC, this is designated  
wabs*zwabs*zpower. 
The values of $z$ were fixed at the values given in Table~1. In cases
where $z$ has not been determined directly, the best estimated
value is used, and if that does not exist (e.g., in the cases of 
GRB980329 and GRB980519) $z$ is allowed to be a free parameter. If all the 
additional red-shifted absorptions are constrained to have the 
same value, the resulting best-fit has a $\chi^2$ of 94.3 for 98 dof
for an average additional red-shifted \nh\ of 
(1.01$\pm ^{0.28}_{0.51}$) 
$\times$ 10$^{22}$ atom cm$^{-2}$. Fixing the red-shifted 
column density to zero and 
re-fitting, results in a $\chi^2$ of 101.1 for 99 dof.
Thus, the addition of this column is significant under an F test 
at the $>$92\% level. 
Since the predicted spectrum in blast wave models varies 
little from event to event (e.g., Wijers et al. 1997),
we next tied the spectral slopes. 
Re-fitting resulted in a $\chi^2$ of 100.4 for 104 dof for an average 
red-shifted \nh\ of (8.7${\pm ^{3.4}_{4.8}}$) 
$\times$ 10$^{21}$ 
atom cm$^{-2}$ and best-fit $\alpha$ of 2.02$\pm ^{0.10}_{0.12}$. 

We note that for strong sources such as the Crab, the 
uncertainty in the fitted \nh\ values is 
a few 10$^{20}$ atom cm$^{-2}$, due 
largely to uncertainties in low energy calibration. 
However, for weak sources, such as afterglows, the 
uncertainty is dominated by the uncertainty 
in background subtraction. From studies 
of weak sources we estimate this to be
$\approxlt$25\% of the statistical error. 

\begin{table*}
\caption[]{GRB afterglow characterstics. The results of absorbed 
power-law model fits are listed. T$\rm _{start}$ is the
time in hrs after the burst that the NFI began observing the afterglow,
T$\rm _{exp}$ is the LECS/MECS exposure time in ks,
OT denotes whether an optical transient is detected, N$\rm _{H}$ is the 
column density in units of 10$^{21}$~atom~cm$^{-2}$, $\alpha$ the 
photon spectral index, N$\rm _{gal}$ the galactic column density, 
and $\beta$ is the power-law temporal decay index of the flux. 
Estimates of the red-shifts, $z$, based on optical 
and logN$-$logF$_{peak}$ data are given. 
All uncertainties are quoted at 68\% confidence}

\begin{tabular}{lccccccccc} 
\hline\noalign{\smallskip}
Burst	& T${ \rm_{start}}$ & T${\rm _{exp}}$ & OT  & N${\rm _{gal}}$	& N${\rm _{H}}$	& $\alpha$	& $\chi$$^{2}$/dof	& $\beta$	& $z$ \\ 
        & (hrs) &  (ks)   &   & &  \\
\noalign{\smallskip\hrule\smallskip}
GRB970228 & 8.0 & 5.5/14.3 & Y	& 1.72	& $2.5 \pm^{1.4}_{1.1}$	& $1.96 \pm^{0.19}_{0.19}$	& 12/10	& 1.33$\pm$0.12 & $\sim$1$^{2}$ \\ 
GRB970402 & 8.0 & 11.6/34.2 & N	& 1.67	& $0.65	\pm^{3.0}_{0.65}$ & $1.67 \pm^{0.32}_{0.57}$	& 6/8	& 1.57$\pm$0.03$^{1}$	& 2.5$^{3}$ \\ 
GRB970508 & 5.7 & 14.1/28.3 & Y	& 0.52	& $1.8	\pm^{2.1}_{1.1}$ & $1.99 \pm^{0.29}_{0.07}$ & 24/22	& 1.1$\pm$0.1& $0.825^{4}$ \\ 
GRB970828 & 28.1 & 36.0 & N	& 0.36	& $3.5	\pm^{2.2}_{1.6}$	& $2.7	\pm^{0.5}_{0.4}$	& 23/31	& 1.44$\pm$0.05	 &\dots \\ 
GRB971214 & 9.7 & 3.2/6.9 & Y	& 0.16	& $4.3	\pm^{6.1}_{2.5}$ & $2.03 \pm^{0.51}_{0.22}$	& 6/8	& 1.20$\pm$0.02& 3.43$^{5}$ \\ 
GRB980329 & 7.0 & 15.8/39.2 & Y	& 0.94	& $10.7 \pm^{7.8}_{4.3}$	& $2.63 \pm^{0.41}_{0.36}$	& 18/27	& 1.35$\pm$0.03	& \dots \\ 
GRB980519 & 9.7 & 23.1/78.2 & Y	& 1.83	& $5.6 \pm^{10.2}_{5.1}$	& $2.52 \pm^{0.70}_{0.57}$ 	& 28/29	& 2.07$\pm$0.11& \dots \\ 
\noalign{\smallskip}
\hline
\mc{10}{l}{\footnotesize $^{1}$Assuming the afterglow 
X-ray emission begins $\sim$100~s after the burst onset;
$^{2}$Wijers et al. (1997);}\\
\mc{10}{l}{\footnotesize  $^{3}$Lipunov et al. (1998); 
$^4$Metzger et al. (1987a);
$^{5}$Kulkarni et al. (1998)}\\
\end{tabular}
\end{table*}

\section{Discussion}

The general properties of afterglows are in remarkable agreement 
with the predictions of even the simplest fireball models (Meszaros \& 
Rees 1997) in which a relativistic blast wave radiates its energy as
it decelerates by plowing through the surrounding medium. As the fireball 
slows down, the peak of the emitted radiation shifts in time to lower 
energies producing the observed power-law decays in the X-ray, optical 
and radio (e.g., Waxman 1997; Wijers \& Galama 1998 
and references therein). The 
diversity in afterglow behavior is most easily explained by beaming 
and/or the differences in radiative losses at early times 
(Meszaros et al. 1998; Wijers \& Galama 1998). However, 
afterglows are only detected optically about half of the time, leading to 
the suggestion that there must be significant extinction at the source. 
The extinction is generally believed to be in the form of dust whose 
existence has been inferred by the observed reddening in the optical 
spectra of GRB971214 and more recently, GRB980329. In addition, the 
detection of the [O~{\sc ii}] 3728\AA\ emission line and also an [Mg~{\sc i}] 
absorption line 
(Metzger et al. 1997a) in GRB970508 
indicates the presence of a relatively dense medium (Metzger et al. 
1997b). Given that the estimated redshifts for 
afterglows lie in the range 1--3.5, 
when star formation was at its peak (e.g., Madau et al. 1998) and that the 
progenitors of GRB970508, GRB971214, GRB980329 and GRB980702 appear to be well 
located inside their 
respective host galaxies, it seems likely that GRB are associated with 
star forming regions. This is supported by the high 
effective temperature implied 
by the relative strengths of the [O~{\sc ii}] and  [Ne~{\sc iii}] 
lines observed in the
optical spectra of GRB970508 and suggests the presence of a substantial 
population of massive stars and thus active star formation
(Bloom et al. 1998). 

The apparent longevity of the observed afterglows and the observation of 
X-ray precursors by {\it Ginga} (Murakami et al. 1991) support a ``dirty'' 
fireball model (Rees \& Meszaros 1998). Although both hypernova and neutron 
star merger models
can satisfy energetics requirements $>$10$^{53}$ ergs s$^{-1}$, 
we would expect a fraction of neutron star mergers to 
take place well outside their host galaxies - purely due to the high proper motion 
acquired in two consecutive supernova explosions.
The fact that the observed optical transients of 4 (and possibly 5) bursts 
are well located within small host galaxies, is suggestive and would tend 
to favor hypernova models.
In these models (Woosley 1993; Paczynski 1998), a very massive 
($\sim$10$\msun$) star collapses, 
producing a ``dirty'' fireball $\sim$300 
times more luminous than a supernova ($\sim$10$^{54}$~erg). This 
large amount 
of energy is obtained from the rotational energy of a Kerr 
black-hole formed in the core collapse. Since such massive stars die 
young ($\sim$10$^6$ yrs) and therefore close to where they were born, 
a natural consequence is 
that GRB trace the star formation rate and should be associated with 
high density, dusty regions. The available data support this. For 
example, by comparing the color spectra of GRB971214 and GRB970508, 
Halpern et al. (1998) calculate 
that an additional \nh\ of (1.5--2.3)$\times$ 10$^{21}$ 
atom cm$^{-2}$ 
is required at the distance of GRB971214 to explain the extreme reddening 
of the spectrum. 
Reichart (1998) has shown that in order to reconcile the optical 
decay and spectral profiles, there must be a red-shifted source of 
extinction at the burst site of 1.9 $\times$ 
10$^{21}$ atom cm$^{-2}$.
He further argues that this absorber is 
probably the host galaxy of GRB971214 at a $z$ of 1.89. 
Taylor et al. (1998) argue from radio observations of 
GRB980329 that an additional \nh\ of 10$^{21}$ atom cm$^{-2}$ is 
required at the source to explain why the afterglow is bright at 
radio wavelengths (e.g., Galama et al. 1998b), but optically dim 
(Palazzi et al. 1998). This 
value is also consistent with that derived from optical 
reddening measurements (Palazzi et al. 1998).


\begin{acknowledgements}
The \sax\ satellite is a joint Italian and Dutch programme. MG,
TO and AO(rr) acknowledge ESA Research Fellowships. We 
thank the staff of the BeppoSAX Science Data Center for help with 
these observations. Peter Meszaros and Jan van Paradijs are thanked
for helpful comments.
\end{acknowledgements}

\end{document}